\documentstyle[11pt]{article}
\begin{document}
\pagestyle{plain}
\huge
\title{\bf From the Foucault pendulum to the galactical gyroscope and LHC}
\large
\author{Miroslav Pardy\\[7mm]
Department of Physical Electronics \\
and\\
Laboratory of Plasma physics\\[5mm]
Masaryk University \\
Kotl\'{a}\v{r}sk\'{a} 2, 611 37 Brno, Czech Republic\\
e-mail:pamir@physics.muni.cz}
\date{\today}
\maketitle
\vspace{5mm}

\begin{abstract}
The Lagrange theory of particle motion in the
noninertial systems is applied to the Foucault
pendulum, isosceles triangle pendulum  and the general
triangle pendulum swinging on the rotating Earth.
As an analogue,  planet orbiting in the rotating
galaxy is considered as the the giant galactical gyroscope.
The Lorentz equation and the Bargmann-Michel-Telegdi
equations are generalized for the rotation system.
The knowledge of these equations  is inevitable for the construction
of LHC where each orbital proton ``feels'' the Coriolis force caused by the
rotation of the Earth.
\end{abstract}

\vspace{3mm}


\vspace{3mm}

\section{Introduction}

In order to reveal the specific characteristics of the mechanical systems in
the rotating framework,
it is necessary to derive the differential equations describing the
mechanical systems in the noninertial systems. We follow the text of
Landau et al. (Landau et al. 1965).

Let be the Lagrange function of a point particle in the inertial
system as follows:

$$L_{0}  = \frac{m{\bf v}_{0}^{2}}{2} - U \eqno(1)$$
with the following equation of motion

$$ m\frac{d{\bf v}_{0}}{dt} = -\frac{\partial U}{\partial {\bf r}},
\eqno(2)$$
where the quantities with index $0$ corresponds to the inertial system.

The Lagrange equations in the noninertial system is of the same form as
that in the inertial one, or,

$$\frac{d}{dt}\frac{\partial L}{\partial {\bf v}} =  \frac{\partial
  L}{\partial {\bf r}}.\eqno(3)$$

However, the Lagrange function in the noninertial system is not the
same as in eq. (1) because it is transformed.

Let us first consider the system $K'$ moving relatively to the system
$K$ with the velocity ${\bf V}(t)$. If we denote the velocity of a
particle with regard to system $K'$ as ${\bf v}'$, then evidently

$${\bf v}_{0} = {\bf v}' + {\bf V}(t).\eqno(4)$$

After insertion of eq. (4) into eq. (1), we get

$$L_{0}'  = \frac{m{\bf v'}^{2}}{2} + m {\bf v'}{\bf V} +
\frac{m}{2}{\bf V }^{2} - U. \eqno(5)$$

The function ${\bf V }^{2}$ is the function of time only and it can be
expressed  as the total derivation of time of some new function. It
means that the term with the total derivation
in the Lagrange function can be removed from the Lagrangian. We also have:

$$ m {\bf v'}{\bf V}(t) = m{\bf V}\frac{d{\bf r'}}{dt} = \frac{d}{dt}
 (m {\bf r'}{\bf V}(t)) - m{\bf r'}\frac{d{\bf V}}{dt}.\eqno(6)$$

After inserting the last formula into the Lagrange function and after
removing the total time derivation we get

$$L' = \frac{mv'^2}{2} - m{\bf W}(t){\bf r'} - U, \eqno(7)$$
where ${\bf W} = d{\bf V}/dt$ is the acceleration the system $K'$.

The Lagrange equations following from the Lagrangian (7) are as
follows:

$$ m\frac{d{\bf v}'}{dt} = -\frac{\partial U}{\partial {\bf r}'}  -
m{\bf W}(t).\eqno(8)$$

We see that after acceleration of the system $K'$ the new force $m{\bf W}(t)$
appears. This force is fictitious one because it is not
generated by the internal properties of some body.

In case that the system $K'$ rotates with the angle
velocity $\bf \Omega$ with regard to the system $K$, the radius
vectors  $\bf r$ and $\bf r'$ are identical and
 (Landau et al., 1965 )

$${\bf v}'  =  {\bf v} + {\bf \Omega}\times{\bf r}.\eqno(9)$$

The Lagrange function for this situation  is  (Landau et al., 1965 )

$$L = \frac{mv^2}{2} - m{\bf W}(t){\bf r} - U + m {\bf v}\cdot({\bf \Omega}
\times{\bf r}) + \frac{m}{2}({\bf \Omega}\times{\bf r})^{2}.\eqno(10)$$

The corresponding  Lagrange equations for the last Lagrange function are as
follows (Landau et al., 1965 ):

$$ m\frac{d{\bf v}}{dt} = -\frac{\partial U}{\partial {\bf r}}  -
m{\bf W} + m{\bf r}\times\dot{\bf \Omega} + 2m{\bf v}\times{\bf \Omega}
+ m{\bf \Omega}\times \left({\bf r}\times{\bf \Omega}\right).\eqno(11)$$

 We observe in eq. (11) three so called inertial forces. The force
$m{\bf r}\times\dot{\bf \Omega}$ is connected with the nonuniform
rotation of the system $K'$ and the forces $2m{\bf v}\times{\bf
\Omega}$ and $m{\bf \Omega}\times {\bf r}\times{\bf \Omega}$
correspond to the uniform rotation. The force $2m{\bf v}\times{\bf\Omega}$
is so called the Coriolis force and it depends on the velocity of a
particle. The force $m{\bf \Omega}\times {\bf r}\times{\bf \Omega}$ is
called the centrifugal force. It is perpendicular to the rotation axes
and the magnitude of it is $m\varrho\omega^{2}$, where $\varrho$ is
the distance of the particle from the rotation axis.

Equation (11) can be applied to many special cases. We apply it first
to the case of the mathematical pendulum swinging
in the gravitational field of the rotating Earth. In other words, to the
so called Foucault pendulum.

\section{Foucault pendulum}

Foucault pendulum was studied by L{\'e}on Foucault (1819 - 1868)
as the big mathematical
pendulum with big mass $m$ swinging in the gravitational field of the Earth. He
used a 67 m long pendulum in the Panth{\'e}on in Paris and showed the
astonished public that the direction of its swing changed over time
rotating slowly.  The experiment proved that the earth rotates. If the
earth would not rotate, the swing would always continue in the same
direction.\footnote{Author performed the experiment with the
Foucault pendulum inside of the rotunda of the Flower garden
in Krom\v e\v r\' i\v z (Moravia, Czech Republic)}

If we consider the motion in the system only with uniform rotation,
then we write equation (11) in the form:

$$ m\frac{d{\bf v}}{dt} = -\frac{\partial U}{\partial {\bf r}}
 + 2m{\bf v}\times{\bf \Omega}
+ m{\bf \Omega}\times {\bf r}\times{\bf \Omega}.\eqno(12)$$

In case of the big pendulum, the vertical motion can be neglected and
at the same time the term with $\Omega^{2}$. The motion of this
pendulum is performed in the horizontal plane xy. The corresponding
equations are as follows (Landau et al., 1965 ):

$$\ddot x + \omega^{2}x = 2\Omega_{z}\dot y, \quad
\ddot y + \omega^{2}y = -2\Omega_{z}\dot x, \eqno(13)$$
where $\omega$ is the frequency of the mathematical pendulum without
rotation of the Earth, or $\omega = 2\pi/T$ and (Landau et al., 1965 ):
$ T \approx  2\pi\sqrt{{l}/{g}}$,
where $T$ is the period of the pendulum oscillations, $l$ is the length
of the pendulum and $g$ is the Earth acceleration.

After multiplication of the second equation of (13) by the imaginary
number $i$ and summation with the first equation, we get:

$$\ddot\xi + 2i\Omega_{z}\dot\xi + \omega^{2}\xi = 0\eqno(14)$$
for the complex quantity $\xi = x + iy$. For the small angle rotation
frequency  $\Omega_{z}$ of the Earth with regard to the oscillation frequency
$\omega$, $\Omega_{z}<< \omega $, we easily find the solution in the form:

$$\xi = e^{-i\Omega_{z}t}(A_{1}e^{i\omega t} + A_{2}e^{-i\omega t} ),
\eqno(15)$$
or,

$$x + iy = e^{-i\Omega_{z}t}(x_{0} + iy_{0}),\eqno(16)$$
where functions $x_{0}(t), y_{0}(t)$ are the parametric expression of
the motion of the pendulum without the Earth rotation. If the complex
number is expressed in the trigonometric form of (16), the
$\Omega_{z}$ is the rotation of the complex number $x_{0} +
iy_{0}$. The physical meaning of eq. (16) is, that the plane of
the Foucault pendulum rotates with the frequency $\Omega_{z}$ with
regard to the Earth.

Galileo Galilei (1564 - 1642) - Italian scientist and philosopher -
studied the mathematical pendulum before Foucault. While in a Pisa
cathedral, he noticed that a chandelier was swinging with the same period
as timed by his pulse, regardless of his amplitude. It is probable, that
Galileo noticed the rotation of the swinging plane of the
pendulum. However, he had not used  this fact as the proof of the Earth
rotation when he was confronted with the Inquizition
tribunal. Nevertheless, his last words were  ``E pur si muove''.

\section{The triangle pendulum}

The triangle pendulum is the analogue of the Foucault pendulum with the
difference that the pendulum is a rigid system composed from a two
rods forming the triangle ABC.
In the isosceles triangle it is $AC = CB = l = const$. The legs
$AC = CB$ are supposed to be prepared from the nonmetal and nonmagnetic
material, with no interaction with the magnetic field of the Earth.
Point $C$ is a vertex at
which the pendulum is hanged. The vertex is realized by the very small
ball. Points $A$ and $B$ are not connected by the rod. The
angle $ACB = \alpha$. The initial deflection angle of $CB$ from the vertical is
$\varphi_{0} + \alpha$, where $\varphi_{0}$ is the initial deflection
angle from vertical.

To be pedagogical clear, let us give first the known
theory of the  simple mathematical pendulum (Amelkin, 1987).

The energetical equation of the pendulum is of the form ($\varphi$ is
the deflection angle from vertical):

$$\frac{mv^2}{2} - mgl\cos\varphi = -mgl\cos\varphi_{0},\eqno(17)$$
from which follows, in the polar coordinates with $v = l\dot\varphi$

$$\ddot\varphi + \frac{g}{l}\sin\varphi = 0. \eqno(18)$$

We have for the very small angle $\varphi$ that $x \approx  l\varphi$
and it means
that from the last equation follows the equation for the harmonic
oscillator

$$\ddot x + \frac{g}{l}x = 0.\eqno(19)$$

The rigorous derivation of the period of pendulum follows from
eq. (17). With $v = ds/dt = l d\varphi/dt$, we get

$$\frac{l}{2}\Bigl(\frac{d\varphi}{dt}\Bigr)^{2} =
g(\cos\varphi - \cos\varphi_{0}).\eqno(20)$$

Then,

$$dt =  \sqrt{\frac{l}{2g}}\frac{d\varphi}
{\sqrt{\cos\varphi - \cos\varphi_{0}}}.\eqno(21)$$

For the period $T$ of the pendulum, we have from the last formula:

$$\frac{T}{4} = \sqrt{\frac{l}{2g}}\int_{0}^{\varphi_{0}}
\frac{d\varphi}{\sqrt{\cos\varphi - \cos\varphi_{0}}}.\eqno(22)$$

Using relations $\cos\varphi = 1 - 2\sin^2\varphi/2,
\cos\varphi_{0} = 1 - 2\sin^2\varphi_{0}/2$,
and substitution $\sin\varphi/2 = k\sin\chi$, with $k =
\sin\varphi_{0}/2$, we get

$$d\varphi = \frac{2\sqrt{k^2 - \sin^2\chi/2}}{\sqrt{1 -
k^2\sin^{2}\chi}}d\chi\eqno(23)$$
and finally

$$T = 4\sqrt{\frac{l}{g}}\int_{0}^{\pi/2}\frac{d\chi}
{\sqrt{1 - k^2\sin^{2}\chi}},\eqno(24)$$
where the integral in the last formula is so called the elliptic
integral, which cannot be evaluated explicitely but only in the form of
series.

Now, let us go back to the isosceles triangle pendulum.
We write in the polar coordinates instead of the equation (17):

$$
\left(\frac{1}{2} ml^2{\dot\varphi}^2 - mgl\cos(\varphi -   \alpha)\right) +
\left(\frac{1}{2} ml^2{\dot\varphi}^2 - mgl\cos\varphi\right) =
const.\eqno(25)$$

Then, after differentiation with regard to time, we get from the last
equation the following one:

$$2\ddot\varphi +  \frac{g}{l}(\sin(\varphi - \alpha) + \sin\varphi) = 0.
\eqno(26)$$

It is easy to see that for $\alpha = 0$ the equation of motion is
$\ddot\varphi + (g/l)\sin\varphi = 0$, which is the expected result
because the triangle pendulum in this case is the mathematical
pendulum.

The equilibrium state of the isosceles triangle pendulum is the state
with $\varphi = \alpha/2$. The small swings are then performed in the
interval

$$\frac{\alpha}{2} - \varepsilon \leq \varphi \leq \frac{\alpha}{2} +
\varepsilon . \eqno(27)$$

If we put

$$\varphi = \chi + \alpha/2 \eqno(28)$$
the equation of motion for the variable $\chi$ is

$$\ddot\chi +  \omega^2\sin\chi = 0 \eqno(29)$$
with

$$\omega = \sqrt{\frac{g}{l}\cos(\alpha/2)}.\eqno(30)$$

For very small angle $\chi$ the equation (29) is the equation of the
harmonic oscillator and in the situation of the rotation of the Earth
it is possible to apply the same mathematical
procedure as
in case of the Foucault pendulum. The result is the same as we have
described it (Landau et al., 1965). In other words, the
triangle pendulum behaves as the Foucault pendulum and it can be used
as the table pendolino experiment  for the demonstration of the Earth rotation.

The triangle pendulum with equal sides can be generalized to the
situation with $AC = l_1$, $BC = l_2$ with masses $m_1, m_2$. Then, it
is easy to show by the same procedure, that the original equation of motion of
such generalized triangle pendulum is as follows:

$$\ddot\varphi + \omega_{1}^{2}\sin(\varphi - \alpha) +
\omega_{2}^{2}\sin(\varphi) = 0, \eqno(31)$$
where

$$\omega_{1}^{2} = \frac{m_{1}gl_{1}}{m_{1}^{2}l_{1}^{2} +
  m_{2}^{2}l_{2}^{2}};\quad
\omega_{2}^{2} = \frac{m_{2}gl_{2}}{m_{1}^{2}l_{1}^{2} +
  m_{2}^{2}l_{2}^{2}}\eqno(32)$$

For $l_{1} = l_{2} = l$ and $m_{1} = m_{2} = m$, we get the isosceles
pendulum and for $\alpha = 0$, we get the original simple mathematical
pendulum.

The mathematical and physical analysis of the general triangle pendulum
shows us that this pendulum has the same behavior as the Foucault
pendulum. Or, in other words we can denote it as the triangle Foucault
pendulum.

Let us still remark that while the magnetic needle of the compass
rotates with the Earth (forced by the magnetic field of the Earth),
the plane of motion of the Foucault pendulum and the triangle pendulum
does not rotate with the Earth.

\section{The galactical gyroscope}

The gyroscope is usually defined as a device for measuring or
maintaining orientation based on the principle of conservation of
angular momentum. The essence of the device is the spinning wheel.
We will show that the planet orbiting in the rotating galaxy is the galactical
gyroscope because the orientation of the orbit is conserved reminding
the classical gyroscope.

The force acting on the planet with mass $m$ is according to Newton law

$$F = - G\frac{m M}{r^{2}},\eqno(33)$$
where $M$ is the mass of Sun, $r$ being the distance from $m$ to the Sun.

The corresponding equations of motion in the coordinate system x and y
are as follows

$$m\ddot x = - G\frac{m M}{r^{2}}\cos \varphi; \quad
m\ddot y = - G\frac{m M}{r^{2}}\sin \varphi,\eqno(34)$$
or, with $\sin\varphi = y/r, \cos\varphi = x/r $,

$$\ddot x = - \frac {kx}{r^{3}}; \quad
\ddot y = - \frac {ky}{r^{3}}, \quad k = GM, \quad r = \sqrt{x^2 +
  y^2} \eqno(35)$$

Using $x = r\cos\varphi, y = r\sin\varphi $, we
get instead of equations (35):

$$(\ddot r - r \dot\varphi^{2})\cos\varphi -
(2\dot r\dot\varphi + r\ddot\varphi){\sin\varphi} =
- \frac{k \cos\varphi}{r^{2}} \eqno(36)$$

$$ (\ddot r - r{\dot\varphi}^{2})\sin\varphi + (2\dot r \dot\varphi +
r\ddot\varphi){\cos\varphi} =
- \frac {k \sin\varphi}{r^{2}}. \eqno(37) $$

In case that the motion of the planet is performed in the rotation
system of a galaxy the equations (36), (37) are written in the form
($\Omega_{z} = \Omega$)

$$(\ddot r -r\dot\varphi^{2})\cos\varphi - (2\dot r\dot\varphi +
r\ddot\varphi)\sin\varphi
 = - \frac {k \cos\varphi}{r^{2}} + 2\Omega\dot y \eqno(38)$$

$$(\ddot r - r\dot\varphi^2)\sin\varphi + (2\dot r \dot\varphi +
r\ddot\varphi)\cos\varphi = - \frac{k \sin\varphi}{r^2} -
2\Omega\dot x ,\eqno(39)$$
or,

$$(\ddot r -r\dot\varphi^2)\cos\varphi - (2\dot r\dot\varphi +
r\ddot\varphi)\sin\varphi
 = - \frac {k\varphi}{r^2} +
2\Omega(\dot r \sin\varphi + r\cos\varphi\dot\varphi)\eqno(40)$$

$$(\ddot r - r\dot\varphi^2)\sin\varphi + (2\dot r \dot\varphi +
r\ddot\varphi)\cos\varphi = - \frac{k\sin\varphi}{r^2}  -
2\Omega(\dot r \cos\varphi - r\sin\varphi\dot\varphi).\eqno(41)$$

After multiplication of eq. (40) by $\sin\varphi $ and eq. (41) by
$\cos\varphi $  and after their subtraction we get

$$2\dot r\dot\varphi + r\ddot\varphi = -2\Omega\dot r,\eqno(42)$$
or,
$$\frac{d}{dt}(r^2\dot\varphi) = - \Omega\frac{d}{dt}(r^2), \eqno(43) $$
or,

$$\dot\varphi = - \Omega .\eqno(44)$$

It means that the angle velocity of the ellipse of a planet inside
the rotating galaxy is $\Omega$ which is the angle velocity of the galaxy.
Let us only remark that here we consider the well defined galaxy as
the galaxy of elliptical form  and not of the chaotic form. We do not
consider here the `` galaxy rotation problem'' - the discrepancy
between the observed  rotation speeds of matter in the disk portion of
spiral galaxies and the predictions of Newton dynamics considering the
luminous mass - which is for instance discussed
 in $http://en.wikipedia.org/wiki/Galaxy_{-}spiral_{-}problem.$

\section{Discussion}

We have presented the Lagrange theory of the noninertial classical
systems and we applied the theory to the so called Foucault pendulum,
the isosceles triangle pendulum  with two equal
masses and to the triangle pendulum with the nonequal legs and masses.
We have shown that Every pendulum is suitable for the demonstration of
the rotation of the earth.

For the demonstration of the galaxy rotation,  we have analyzed the
elliptical motion of our planet and we have shown that the orbital
motion of our planet can be used as gigantic
gyroscope for the proof of the rotation of our
galaxy in the universe. The orbit of our planet with regard to the rest
of the universe has the stable stationary position  while the galaxy
rotates. The orbital planetary stability
can be used as the method of the  investigation  of
the rotation of all galaxies in the
rest of the universe. To our knowledge this method was not still used in
the galaxy astrophysics.

It is possible to consider also the rotation of the
Universe. If we define Universe as the material bodies immersed into
vacuum, then the rotation of the Universe is physically meaningful
and the orbit of our planet is of the constant position with regard to
the vacuum
as the rest system. The idea that the vacuum is the rest system is
physically meaningful because only vacuum is the origin of the inertial
properties of every massive body. In other words, the inertial mass
$m$ in the Newton-Euler  equation $F = ma$ is the result of the interaction of
the massive body with vacuum and in no case it is the result of the Mach
principle where the inertial mass is generated  by the mass of rest of the
Universe. At present time, everybody knows that Mach principle is
absolutely invalid for all time of the existence of Universe.

Now, the question arises what is the description of the rotation in
the general theory of relativity. If we use the the Minkowski element

$$ds^2  =  c^2{dt'}^2  - {dx'}^2 -  {dy'}^2 - {dz'}^2 \eqno(45)$$
and the nonrelativistic transformation to the rotation system (Landau
et al., 1988)

$$x' = x\cos\Omega t - y\sin\Omega t , \quad y' = x\sin\Omega t +
y\cos\Omega t ,\quad z = z'\eqno(46)$$
then we get:

$$ds^2  =  [c^2 - \Omega^2 (x^2 + y^2)]{dt}^2  - {dx}^2 -  {dy}^2 -
{dz}^2  + 2\Omega y dxdt - 2\Omega x dydt, \eqno(47)$$
which is not relativistically invariant.

If we use the Minkowski element in the cylindrical coordinates

$$ds^2  =  c^2{dt'}^2  - {dr'}^2 - {r'}^2 {d\varphi'}^2 - {dz'}^2 \eqno(48)$$
and the transformation to the rotating system $r' = r, z' = z,
\varphi' = \Omega t$ (Landau et all., 1988), we get the noninvariant
element

$$ds^2  =  [c^2 - \Omega^2 r^2]{dt}^2  - 2\Omega r^2 d\varphi dt -
{dr}^2 - {r}^2 {d\varphi}^2 - {dz}^2. \eqno(49)$$

So, we see that the rotational system  can be used only for for $r <
c/{\Omega}$. For $r >  c/{\Omega}$, the component $g_{00}$ is negative,
which is in the contradiction with the principles of relativity.

However, according to special theory of relativity only Lorentz transformation
can be inserted into the equation (45). Or,

$$dx' = \gamma(dx - vdt), \quad dt' = \gamma(t - vx/c^2),
\quad \gamma = \frac{1}{\sqrt{} 1 - v^2/c^2}.\eqno(50)$$

The radial coordinate of the rotational system is not contracted.
Only the tangential coordinate $dx$. So, in the cylindrical coordinates,
it is necessary to write $dx = rd\varphi,dx' = rd\varphi', v =
\Omega r$. Using these ingredients we write:

$$d\varphi' = \gamma(d\varphi - \Omega dt), \quad dt' = \gamma(dt -
(\Omega r^2/c^2)d\varphi).\eqno(51)$$

After insertion of equations (51) into equation (48), we get for the
interval $ds$ the original relation (48):

$$ds^2 = c^2{dt}^2 - {dr}^2  - r^2{d\varphi}^2 - {dz}^2, \eqno(52)$$

We think that transformation (51) is correct because it it based on the Lorentz
transformation, which has here the physical meaning of the relation
between tangential elements $dx = rd\varphi, dx' = rd\varphi', v = \Omega r$
and the infinitesimal time relation.

The correctness of the transformation between inertial and rotation
system is necessary because it enables
to describe the motion of the particle and spin in the LHC by the
general relativistic methods. The basic idea is the generalization so
called Lorentz equation for the charged particle in the
electromagnetic field $F^{\mu\nu}$ (Landau et all., 1988):

$$mc \frac{dv^{\mu}}{ds} = \frac{e}{c}F^{\mu\nu}v_{\nu}.\eqno(53)$$

In other words, the normal derivative is replaced by the covariant one and
we get the general relativistic equation for the motion of a charged
particle in the electromagnetic field and gravity (Landau et all.,
1988):

$$ mc \left(\frac{dv^{\mu}}{ds} +
 \Gamma^{\mu}_{\alpha\beta}v^{\alpha}v^{\beta}\right)
 = \frac{e}{c}F^{\mu\nu}v_{\nu}, \eqno(54)$$
where

$$\Gamma^{\mu}_{\alpha\beta} = \frac{1}{2}g^{\mu\lambda}
\left(\frac{\partial g_{\lambda\alpha}}{\partial x^{\beta}} +
\frac{\partial g_{\lambda\beta}}{\partial x^{\alpha}} -
\frac{\partial g_{\alpha\beta}}{\partial x^{\lambda}}
\right)\eqno(55)$$
are the Christoffel symbols derived in the Riemann geometry theory
(Landau et all., 1988).

In case that we consider motion in the
rotating system, then it is necessary to insert the metrical
tensor $g_{\mu\nu}$, following from the Minkowski element for the
rotation system. The construction of LHC with proton  must
be in harmony with equation (54) because orbital protons ``feels'' the Coriolis force
from the rotation of the Earth.

The analogical situation occurs for the motion of the spin. While the
original Bargmann-Michel-Telegdi equation for the spin motion is as
follows (Berestetzkii et all., 1988)

$$ \frac{da^{\mu}}{ds} = 2\mu F^{\mu\nu}a_{\nu} - 2\mu'\mu v^{\mu}
F^{\alpha\beta}v_{\alpha}a_{\beta}, \eqno(56)$$
where $\mu' = \mu - e/2m$ and $a_{\mu}$ is the axial vector, which
follows also from the classical limit of the Dirac equation as
$\bar\psi i\gamma_{5}\gamma_{\mu}\psi$ (Rafanelli et all., 1964; Pardy, 1973),
the general relativistic generalization of the Bargmann-Michel-Telegdi
equation  can be obtained by the analogical procedure which was
performed with the Lorentz equation. Or,

$$ \left(\frac{da^{\mu}}{ds} +
  \Gamma^{\mu}_{\alpha\beta}v^{\alpha}a^{\beta}\right)
= 2\mu F^{\mu\nu}a_{\nu} - 2\mu'\mu v^{\mu}
F^{\alpha\beta}v_{\alpha}a_{\beta}, \eqno(57)$$
where in case of the rotating system the metrical tensor $g_{\mu\nu}$
must be replaced by the metrical tensor of the rotating system. Then,
the last equation will describe the motion of the spin in the rotating
system.

The motion of the polarized proton in LHC will be described by the last
equation because our Earth rotates. During the derivation we wrote
$\Gamma^{\mu}_{\alpha\beta}v^{\alpha}a^{\beta}$
and not $\Gamma^{\mu}_{\alpha\beta}v^{\alpha}v^{\beta}$, because every
term must be the axial vector. In other words, the last
equation for the motion of the spin in the rotating system was not
strictly derived but created with regard to the philosophy that
physics is based on the creativity and logic (Pardy, 2005).

On the other hand, the equation (57) must evidently follow from the
Dirac equation in the rotating system, by the same WKB methods which
were  used by  Rafanelli, Schiller and Pardy
( Rafanelli and  Schiller, 1964; Pardy, 1973).
The derived BMT equation in the metric  of the rotation of the Earth
are fundamental for the proper work of LHC because every orbital
proton of  LHC
``feels'' the rotation of the Earth and every orbital proton spin ``feels''
the Earth rotation too. So, LHC needs equations (54) and (57) and vice versa.

The theory discussed in our article can be also applied to the
pendulum where the fibre is elastic. The corresponding motion is
then described by the wave equation with the initial and boundary
conditions.

It is evident that there are many physical problems, classical and
quantum mechanical considered in the rotation system. Some problems were solved
and some problems will be solved in the future. Let us define some of
these problems.

M\"ossbauer effect in the rotating system, Schr\"odinger equation  for
a particle in the rotating system,  Schr\"odinger equation  for the
pendulum in the inertial system and in the rotating system,  Schr\"odinger
equation of H-atom  in the rotating system,  Schr\"odinger equation of
harmonic  oscillator in the rotating system, the \v Cerenkov effect in the
rotating dielectric medium , the relic radiation in the rotating galaxy,
The N-dimensional blackbody radiation in the rotating system,
conductivity and superconductivity in the
rotating system, laser pulse in the rotating system, Berry phase,
Sagnac effect,
and so on. All these problems can be formulated classically, or in the
framework of the general theory of relativity with the $\Gamma$-connections
corresponding to the geometry of the rotating system.
We hope that the named problems are interesting and their
solution will be integral part of the theoretical physics.

\vspace{15mm}

{\bf References}

\vspace{5mm}

\noindent
Amelkin, V. V. (1987). Applied Differential equations, (Moscow,
Nauka), (in Russian).\\[2mm]
Berestetzkii, V. B., Lifshitz, E. M. and Pitaevskii, L. P. (1999).
Quantum electrodynamics, (Butterworth-Heinemann, Oxford).\\[2mm]
Landau, L. D. and Lifshitz, E. M. (1965). Mechanics, (Moscow,
Nauka), (in Russian). \\[2mm]
Landau, L. D. and Lifshitz, E. M. (2000). The Classical Theory of
Fields, (Butterworth-Heinemann, Oxford). \\[2mm]
Rafanelli, K, and Schiller, R. (1964). Classical motion of
spin-1/2 particles, {\it Phys. Rev.} {\bf 135}, No. 1 B, B279.\\[2mm]
Pardy, M. (1973). Classical motion of spin 1/2 particles with zero
anomalous magnetic moment, {\it Acta Physica Slovaca} {\bf 23},
No. 1, 5. \\[2mm]
Pardy, M. (2005). Creativity leading to discoveries in particle
physics and relativity, physics/0509184.

\end{document}